\documentclass[english,prd,english,notitlepage,nofootinbib]{revtex4}
\usepackage[latin9]{inputenc}
\usepackage{babel}
\usepackage{amsmath}
\usepackage{amssymb}
\usepackage{graphicx}
\usepackage{esint}
\usepackage[unicode=true,pdfusetitle,
 bookmarks=true,bookmarksnumbered=false,bookmarksopen=false,
 breaklinks=false,pdfborder={0 0 1},backref=false,colorlinks=false]
 {hyperref}

\makeatletter

\providecommand{\tabularnewline}{\\}

\@ifundefined{textcolor}{}
{%
 \definecolor{BLACK}{gray}{0}
 \definecolor{WHITE}{gray}{1}
 \definecolor{RED}{rgb}{1,0,0}
 \definecolor{GREEN}{rgb}{0,1,0}
 \definecolor{BLUE}{rgb}{0,0,1}
 \definecolor{CYAN}{cmyk}{1,0,0,0}
 \definecolor{MAGENTA}{cmyk}{0,1,0,0}
 \definecolor{YELLOW}{cmyk}{0,0,1,0}
}

\usepackage{babel}

\makeatother

\begin{document}

\title{Loop corrections to pion and kaon neutrinoproduction}

\author{Marat~Siddikov, Iv\'an~Schmidt}

\address{Departamento de Física, Universidad Técnica Federico Santa María,\\
 y Centro Científico - Tecnológico de Valparaíso, Casilla 110-V, Valparaíso,
Chile}

\begin{abstract}
In this paper we study the next-to-leading order corrections to 
deeply virtual pion and kaon production in neutrino experiments. We
estimate these corrections in the kinematics of the \textsc{Minerva} experiment
at FERMILAB, and find that they are sizable and increase the leading
order cross-section by up to a factor of two. We provide a computational
code, which can be used for the evaluation of the cross-sections, taking into
account these corrections and employing various GPD models. 
\end{abstract}



\maketitle

\section{Introduction}

Today generalized parton distributions (GPDs) are used as a common
language to parametrize the nonperturbative structure of the target.
In Bjorken kinematics, due to collinear factorization theorems~\cite{Ji:1998xh,Collins:1998be},
these objects can be accessed in a study of cross-sections for a wide
class of processes. Nowadays all the information on GPDs comes from
the electron-proton and positron-proton measurements done at JLAB
and HERA, in particular deeply virtual Compton scattering (DVCS) and
deeply virtual meson production (DVMP)~\cite{Ji:1998xh,Collins:1998be,Mueller:1998fv,Ji:1996nm,Ji:1998pc,Radyushkin:1996nd,Radyushkin:1997ki,Radyushkin:2000uy,Collins:1996fb,Brodsky:1994kf,Goeke:2001tz,Diehl:2000xz,Belitsky:2001ns,Diehl:2003ny,Belitsky:2005qn,Kubarovsky:2011zz}.
However, due to the rich structure of GPDs, as well as the limited available
experimental data, modern parametrizations of GPDs still rely significantly
on various additional assumptions. A planned CLAS12 upgrade at JLAB
will improve our understanding of GPDs and will extend the kinematic
coverage~\cite{Kubarovsky:2011zz}. Nevertheless, the flavor structure of
GPDs has been experimentally tested only for limiting cases, PDFs
and form factors, and its extrapolation to a broader kinematical range
still rests on additional implicit assumptions. This happens because
DVCS alone only probes a certain flavor combination, whereas analysis
of \emph{e}DVMP is aggravated by large uncertainties from twist-3
effects for pion production~\cite{Kubarovsky:2011zz,Ahmad:2008hp,Goloskokov:2009ia,Goloskokov:2011rd,Goldstein:2012az}
and from lack of knowledge of meson distribution amplitudes (DAs) for vector meson production.

Earlier~\cite{Kopeliovich:2012dr} we suggested that GPDs could be
studied in neutrino-induced deeply virtual meson production ($\nu$DVMP)
of the pseudo-Goldstone mesons ($\pi,\, K,\,\eta$), using the high-intensity
\textsc{NuMI} beam at Fermilab. Right now it runs in the so-called middle-energy
(ME) regime, with an average neutrino energy of about 6~GeV, although
without major rebuild potentially it could deliver neutrinos with energies
up to 20 GeV~\cite{Drakoulakos:2004gn}. For studies of flavor structure,
$\nu$DVMP has a clear advantage compared to \emph{electro}production:
since in addition to the vector channel, which is sensitive to a small helicity
flip GPDs $\tilde{H}$ and $\tilde{E}$, and easily gets shadowed
by twist-3 effects~\cite{Kubarovsky:2011zz} at moderate virtualities
$Q^{2}$, the axial part of the weak current gets large contributions
from the unpolarized GPDs, $H,\, E$. As we have shown earlier~\cite{Kopeliovich:2014pea},
in the case of $\nu$DVMP, due to these contributions, the admixture of twist-3
terms becomes negligible~%
\footnote{In this respect we differ from~\cite{Goldstein:2009in}, where it
was assumed that the twist-3 contribution is the dominant mechanism.%
}. An additional appeal of the axial channel stems from the closeness
of pion and kaon distribution amplitudes, guaranteed by chiral
symmetry breaking: neglecting the difference of masses of the two
goldstones in Bjorken kinematics, we may consider the final state mesons
as natural filters of the different GPDs flavor combinations. A suppression
of Cabibbo forbidden, strangeness changing processes can be avoided
if kaon production is accompanied by the conversion of a nucleon to strange
baryons $\Lambda$ and $\Sigma^{\pm,0}$: in such processes the transition
GPDs are related by $SU(3)$ relations \cite{Frankfurt:1999fp} to
linear combinations of different flavor components of the nucleon GPDs.
If all the suggested channels will be measured by MINERvA, a full
light flavor structure of nucleon GPDs could be extracted. Recently
it was suggested in~\cite{Pire:2015iza,Pire:2015vxa,Pire:2016jtr}
that this approach could be extended to $D$-meson production, a challenge
for future high-energy neutrino experiments. However, that kinematics
analysis is more complicated due to additional Bethe-Heitler type
corrections~\cite{Kopeliovich:2013ae}, which are small in MINERvA
kinematics, but grow with $Q^{2}$ and eventually become the dominant
mechanism. Moreover, the fact that nuclear targets are used in neutrino experiments
does not introduce significant uncertainty if we consider incoherent
scattering at sufficiently large recoil momenta $|t|\gg R_{A}^{-2}$,
where $R_{A}$ is the nuclear radius~\cite{Schmidt:2015nka}.

Given the potential of neutrinoproduction of goldstone mesons for
proton structure studies, in this paper we proceed with the corresponding analysis
and evaluate the next-to-leading order (NLO) corrections to DVMP.
As we will show below, for the kinematics of ongoing and forthcoming
neutrino experiments, where typical virtualities $Q^{2}$ are not
very large, the NLO corrections are significant and affect the analysis
of the DVMP contributions. The paper is organized as follows. In Section~\ref{sec:DVMP_Xsec}
we evaluate the goldstone meson production by neutrinos on nucleon
targets, taking into account higher twist effects. In Section~\ref{sec:Parametrizations},
for the sake of completeness, we sketch the properties of the GPDs parametrization
which will be used for evaluations. In Section~\ref{sec:Results}
we present numerical results and conclusions.

\section{Cross-section of the $\nu$DVMP process}

\label{sec:DVMP_Xsec}The cross-section of goldstone mesons production
in neutrino-hadron collisions has a form
\begin{align}
\frac{d\sigma}{dt\, dx_{B}dQ^{2}} & =\Gamma\sum_{\nu\nu'}\mathcal{A}_{\nu',\nu L}^{*}\mathcal{A}_{\nu',\nu L},\label{eq:sigma_def}
\end{align}
where $t=\left(p_{2}-p_{1}\right)^{2}$ is the momentum transfer to the
baryon, $Q^{2}=-q^{2}$ is the virtuality of the charged boson, $x_{B}=Q^{2}/(2p\cdot q)$
is the Bjorken variable, the subscript indices $\nu$ and $\nu'$ in the amplitude
$\mathcal{A}$ refer to helicity states of the baryon before and after
interaction, and the letter $L$ reflects the fact that in the Bjorken
limit the dominant contribution comes from the longitudinally polarized
massive bosons $W^{\pm}/Z$~\cite{Ji:1998xh,Collins:1998be}. The
kinematic factor $\Gamma$ in~(\ref{eq:sigma_def}) is different
for charged current and neutral current processes and is given explicitly by
\begin{align}
\Gamma_{CC} & =\frac{G_{F}^{2}f_{M}^{2}x_{B}^{2}\left(1-y-\frac{\gamma^{2}y^{2}}{4}\right)}{64\pi^{3}Q^{2}\left(1+Q^{2}/M_{W}^{2}\right)^{2}\left(1+\gamma^{2}\right)^{3/2}},\\
\Gamma_{NC} & =\frac{G_{F}^{2}f_{M}^{2}x_{B}^{2}\left(1-y-\frac{\gamma^{2}y^{2}}{4}\right)}{64\pi^{3}\cos^{4}\theta_{W}Q^{2}\left(1+Q^{2}/M_{Z}^{2}\right)^{2}\left(1+\gamma^{2}\right)^{3/2}}.
\end{align}
where $\theta_{W}$ is the Weinberg angle, $M_{W}$ and $M_{Z}$ are
the masses of the heavy bosons $W^{\pm}$ and $M_{Z}$, $G_{F}$ is
the Fermi constant, $f_{M}$ is the produced meson (pion or kaon)
decay constant, and we have introduced the shorthand notations 
\begin{equation}
\gamma=\frac{2\, m_{N}x_{B}}{Q},\quad y=\frac{Q^{2}}{s_{\nu p}\, x_{B}}=\frac{Q^{2}}{2m_{N}E_{\nu}\, x_{B}}.\label{eq:elasticity}
\end{equation}
where $E_{\nu}$ is the neutrino energy in the target rest frame.

\begin{figure}[htp]
\includegraphics[scale=0.75]{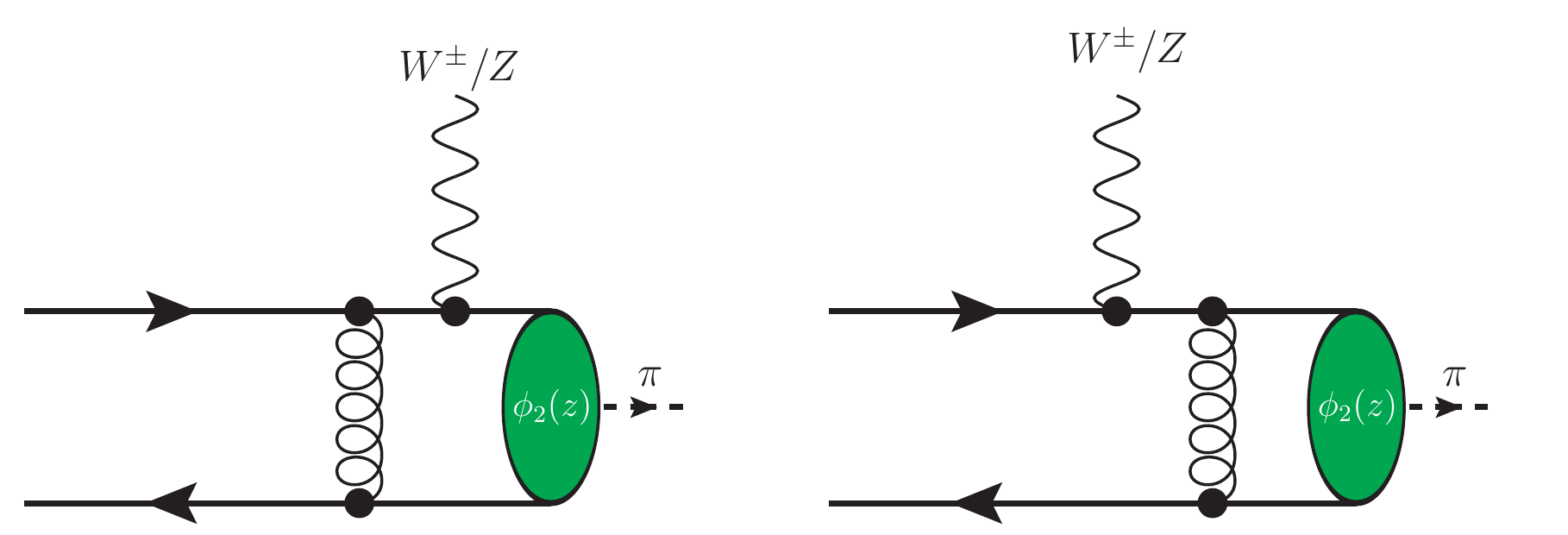} \protect\caption{\label{fig:DVMPLO}Leading-order contributions to the DVMP hard coefficient
functions. Green blob stands for the pion wave function. Additional
diagrams (not shown) may be obtained reversing directions of the quark
lines. }
\end{figure}

\begin{figure}[htp]
\includegraphics[scale=0.85]{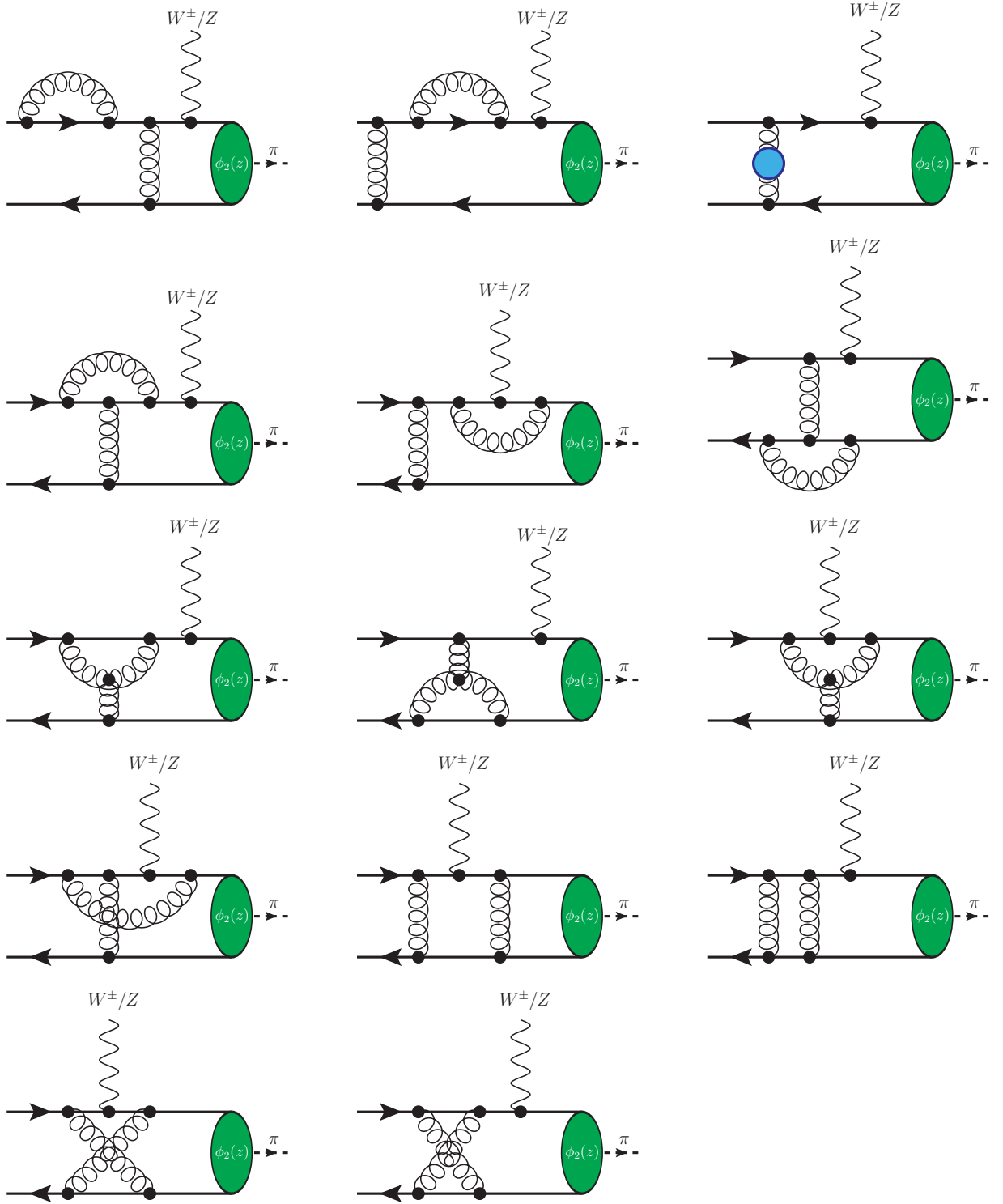} \protect\caption{\label{fig:DVMPNLO}Next-to-leading-order contributions to the DVMP
hard coefficient functions. Green blob stands for the pion wave function.
Blue circle in the third diagram in the first line stands for all
possible gluon mass corrections (sum of quark and gluon loops). Additional
diagrams (not shown) may be obtained reversing directions of the quark
lines.}
\end{figure}

Thanks to the factorization theorem, the amplitude~$\mathcal{A}_{\nu',\nu L}$
in~(\ref{eq:sigma_def}) may be written as a convolution of hard
and soft parts,
\begin{equation}
\mathcal{A}_{\nu',\nu L}=\int_{-1}^{+1}dx\sum_{q,q'=u,d,s,g}\sum_{\lambda\lambda'}\mathcal{H}_{\nu'\lambda',\nu\lambda}^{q'q}\mathcal{C}_{\lambda',\lambda L}^{q'q},\label{eq:M_conv}
\end{equation}
where $x$ is the average light-cone fraction of the parton, $\lambda,\, q$
($\lambda',\, q'$) are the corresponding helicity and flavor of the
initial (final) partons, and $\mathcal{C}_{\lambda'\nu',\lambda\nu}^{q}$
is the hard coefficient function, which will be specified later. The
soft matrix element $\mathcal{H}_{\nu'\lambda',\nu\lambda}^{q}$ in~(\ref{eq:M_conv_2})
is diagonal in quark helicities ($\lambda,\lambda'$), at leading twist,
\begin{align}
\mathcal{H}_{\nu'\lambda',\nu\lambda}^{q'q} & =\frac{2\delta_{\lambda\lambda'}}{\sqrt{1-\xi^{2}}}\left(-g_{A}^{q}\left(\begin{array}{cc}
\left(1-\xi^{2}\right)H^{q'q}-\xi^{2}E^{q'q} & \frac{\left(\Delta_{1}+i\Delta_{2}\right)E^{q'q}}{2m}\\
-\frac{\left(\Delta_{1}-i\Delta_{2}\right)E^{q'q}}{2m} & \left(1-\xi^{2}\right)H^{q'q}-\xi^{2}E^{q'q}
\end{array}\right)_{\nu'\nu}\right.\\
 & +\left.{\rm sgn}(\lambda)g_{V}^{q}\left(\begin{array}{cc}
-\left(1-\xi^{2}\right)\tilde{H}^{q'q}+\xi^{2}\tilde{E}^{q'q} & \frac{\left(\Delta_{1}+i\Delta_{2}\right)\xi\tilde{E}^{q'q}}{2m}\\
\frac{\left(\Delta_{1}-i\Delta_{2}\right)\xi\tilde{E}^{q'q}}{2m} & \left(1-\xi^{2}\right)\tilde{H}^{q'q}-\xi^{2}\tilde{E}^{q'q}
\end{array}\right)_{\nu'\nu}\right),\nonumber 
\end{align}
where the constants $g_{V}^{q},\, g_{A}^{q}$ are the vector and axial
current couplings to quarks, and the four leading twist GPDs $H^{q'q},\, E^{q'q},\,\tilde{H}^{q'q}$
and $\tilde{E}^{q'q}$ are defined as 
\begin{eqnarray}
\frac{\bar{P}^{+}}{2\pi}\int dz\, e^{ix\bar{P}^{+}z}\left\langle B\left(p_{2}\right)\left|\bar{\psi}_{q'}\left(-\frac{z}{2}\right)\gamma_{+}\psi_{q}\left(\frac{z}{2}\right)\right|A\left(p_{1}\right)\right\rangle  & = & \left(H^{q'q}\left(x,\xi,t\right)\bar{N}\left(p_{2}\right)\gamma_{+}N\left(p_{1}\right)\right.\label{eq:H_def}\\
 &  & \left.+\frac{\Delta_{k}}{2m_{N}}E^{q'q}\left(x,\xi,t\right)\bar{N}\left(p_{2}\right)i\sigma_{+k}N\left(p_{1}\right)\right)\nonumber \\
\frac{\bar{P}^{+}}{2\pi}\int dz\, e^{ix\bar{P}^{+}z}\left\langle B\left(p_{2}\right)\left|\bar{\psi}_{q'}\left(-\frac{z}{2}\right)\gamma_{+}\gamma_{5}\psi_{q}\left(\frac{z}{2}\right)\right|A\left(p_{1}\right)\right\rangle  & = & \left(\tilde{H}^{q'q}\left(x,\xi,t\right)\bar{N}\left(p_{2}\right)\gamma_{+}\gamma_{5}N\left(p_{1}\right)\right.\label{eq:Htilde_def}\\
 &  & \left.+\frac{\Delta_{+}}{2m_{N}}\tilde{E}^{q'q}\left(x,\xi,t\right)\bar{N}\left(p_{2}\right)N\left(p_{1}\right)\right),\nonumber 
\end{eqnarray}
with $\bar{P}=p_{1}+p_{2}$, $\Delta=p_{2}-p_{1}$ and $\xi=-\Delta^{+}/2\bar{P}^{+}\approx x_{Bj}/(2-x_{Bj})$
(see e.g.~\cite{Goeke:2001tz} for details of the kinematics). In the
case when the baryon state does not change, $A=B$, the corresponding
GPDs are diagonal in flavor space, $H^{q'q}\sim\delta_{q'q}H^{q}$,
etc. In the general case, when $A\not=B$, in the right-hand side
(r.h.s.) of Eqs.~(\ref{eq:H_def}), (\ref{eq:Htilde_def}) there
might be extra structures, which vanish due to $T$-parity in the case
$A=B$~\cite{Goeke:2001tz}. In what follows we assume that the
initial target $A$ is either a proton or a neutron, and $B$ belongs
to the same lowest $SU(3)$ octet of baryons. In this case, all such
terms are parametrically suppressed by the current quark mass $m_{q}$
and vanish in the limit of exact $SU(3)$, so we will disregard them.
In this special case, we may use $SU(3)$ relations and express the
nondiagonal transitional GPDs as linear combinations of the GPDs of
the proton $H^{q},\, E^{q},\,\tilde{H}^{q},\,\tilde{E}^{q}$~\cite{Frankfurt:1999fp},
so~(\ref{eq:M_conv}) may be effectively rewritten as 
\begin{equation}
\mathcal{A}_{\nu'0,\nu\alpha}=\int_{-1}^{+1}dx\sum_{q=u,d,s}\sum_{\lambda\lambda'}\mathcal{H}_{\nu',\nu}^{q}\mathcal{C}^{q}.\label{eq:M_conv_2}
\end{equation}
where $\mathcal{C}^{q}$ is the helicity independent part of the hard
coefficient function. Its evaluation is quite straightforward, and
in the leading order over $\alpha_{s}$ it gets contributions from the
diagrams shown schematically in Figure~\ref{fig:DVMPLO}. It
has been studied both for pion electroproduction~\cite{Vanderhaeghen:1998uc,Mankiewicz:1998kg,Goloskokov:2006hr,Goloskokov:2007nt,Goloskokov:2008ib,Goloskokov:2011rd,Goldstein:2012az}
and neutrinoproduction~\cite{Kopeliovich:2012dr} (see also~\cite{Kopeliovich:2013ae}
for a discussion of higher twist corrections). In the next-to-leading
order the evaluation becomes more complicated, and the corresponding diagrams
are shown schematically in Figure~\ref{fig:DVMPNLO}.

Straightforward evaluation of the diagrams shown in the Figure~\ref{fig:DVMPLO}
yields for the coefficient function
\begin{align}
\mathcal{C}_{\lambda',\lambda\mu}^{q} & =\eta_{-}^{q}c_{-}^{(2)}\left(x,\xi\right)+{\rm sgn}(\lambda)\eta_{+}^{q}c_{+}^{(2)}\left(x,\xi\right)+\mathcal{O}\left(\frac{m^{2}}{Q^{2}}\right)+\mathcal{O}\left(\alpha_{s}^{2}\left(\mu_{R}^{2}\right)\right)\label{eq:Coef_function}
\end{align}
where the process-dependent flavor factors $\eta_{V\pm}^{q},\,\eta_{A\pm}^{q}$
are given in Table~\ref{tab:DVMP_amps}~%
\footnote{As was discussed above, for processes with change of internal
baryon structure, we use $SU(3)$ relations~\cite{Frankfurt:1999fp}
which are valid up to corrections in current quark masses $\sim\mathcal{O}\left(m_{q}\right)$.%
}, and we introduced shorthand notations

\begin{table}[h]
\protect\caption{\label{tab:DVMP_amps}The flavor coefficients $\eta_{\pm}^{q}$ for
several pion and kaon production processes discussed in this paper
($q=u,d,s,...$). For the case of CC mediated processes, take $\eta_{V\pm}^{q}=\eta_{\pm}^{q},\quad\eta_{A\pm}^{q}=-\eta_{\pm}^{q}$.
For the case of NC mediated processes, take $g_{q}$ corresponding
to $g_{V}^{q}$ and $g_{A}^{q}$ for helicity odd and helicity even
GPDs respectively.}

\global\long\def\arraystretch{1.5}

\begin{tabular}{|c|c|c|c|c|c|c|c|c|}
\cline{1-4} \cline{6-9} 
Process  & type  & $\eta_{+}^{q}$ & $\eta_{-}^{q}$ &  & Process  & type  & $\eta_{+}^{q}$ & $\eta_{-}^{q}$\tabularnewline
\cline{1-4} \cline{6-9} 
$\nu\, p\to\mu^{-}\pi^{+}p$  & CC  & $V_{ud}\delta_{qu}$ & $V_{ud}\delta_{qd}$ &  & $\nu\, n\to\mu^{-}\pi^{+}n$  & CC  & $V_{ud}\delta_{qd}$ & $V_{ud}\delta_{qu}$\tabularnewline
\cline{1-4} \cline{6-9} 
$\bar{\nu}\, p\to\mu^{+}\pi^{-}p$  & CC  & $V_{ud}\delta_{qd}$ & $V_{ud}\delta_{qu}$ &  & $\bar{\nu}\, n\to\mu^{+}\pi^{-}n$  & CC  & $V_{ud}\delta_{qu}$ & $V_{ud}\delta_{qd}$\tabularnewline
\cline{1-4} \cline{6-9} 
$\bar{\nu\,}p\to\mu^{+}\pi^{0}n$  & CC  & $V_{ud}\frac{\delta_{qu}-\delta_{qd}}{\sqrt{2}}$ & $-V_{ud}\frac{\delta_{qu}-\delta_{qd}}{\sqrt{2}}$ &  & $\nu\, n\to\mu^{-}\pi^{0}p$  & CC  & $-V_{ud}\frac{\delta_{qu}-\delta_{qd}}{\sqrt{2}}$ & $V_{ud}\frac{\delta_{qu}-\delta_{qd}}{\sqrt{2}}$\tabularnewline
\cline{1-4} \cline{6-9} 
$\nu\, p\to\nu\,\pi^{+}n$  & NC  & $g_{d}\left(\delta_{qu}-\delta_{qd}\right)$ & $g_{u}\left(\delta_{qu}-\delta_{qd}\right)$ &  & $\nu\, n\to\nu\,\pi^{-}p$  & NC  & $g_{u}\left(\delta_{qu}-\delta_{qd}\right)$ & $g_{d}\left(\delta_{qu}-\delta_{qd}\right)$\tabularnewline
\cline{1-4} \cline{6-9} 
$\nu\, p\to\nu\,\pi^{0}p$  & NC  & $\frac{g_{u}\delta_{qu}-g_{d}\delta_{qd}}{\sqrt{2}}$ & $\frac{g_{u}\delta_{qu}-g_{d}\delta_{qd}}{\sqrt{2}}$ &  & $\nu\, n\to\nu\,\pi^{0}n$  & NC  & $\frac{g_{u}\delta_{qd}-g_{d}\delta_{qu}}{\sqrt{2}}$ & $\frac{g_{u}\delta_{qd}-g_{d}\delta_{qu}}{\sqrt{2}}$\tabularnewline
\cline{1-4} \cline{6-9} 
\multicolumn{1}{c}{} & \multicolumn{1}{c}{} & \multicolumn{1}{c}{} & \multicolumn{1}{c}{} & \multicolumn{1}{c}{} & \multicolumn{1}{c}{} & \multicolumn{1}{c}{} & \multicolumn{1}{c}{} & \multicolumn{1}{c}{}\tabularnewline
\cline{1-4} \cline{6-9} 
$\nu\, p\to\mu^{-}K^{+}p$  & CC  & $V_{us}\delta_{qu}$ & $V_{us}\delta_{qs}$ &  & $\nu\, n\to\nu\, K^{+}\Sigma^{-}$  & NC  & $-g_{d}\left(\delta_{qu}-\delta_{qs}\right)$ & $-g_{u}\left(\delta_{qu}-\delta_{qs}\right)$\tabularnewline
\cline{1-4} \cline{6-9} 
$\nu\, p\to\mu^{-}K^{+}\Sigma^{+}$  & CC  & 0 & $-V_{ud}\left(\delta_{qd}-\delta_{qs}\right)$ &  & $\nu\, n\to\mu^{-}K^{+}\Sigma^{0}$  & CC  & 0 & $-V_{ud}\frac{\delta_{qu}-\delta_{qs}}{\sqrt{2}}$\tabularnewline
\cline{1-4} \cline{6-9} 
$\bar{\nu}\, n\to\mu^{+}K^{0}\Sigma^{-}$  & CC  & 0 & $-V_{ud}\left(\delta_{qu}-\delta_{qs}\right)$ &  & $\nu\, p\to\nu\, K^{+}\Lambda$  & NC  & $-g_{d}\frac{2\delta_{qu}-\delta_{qd}-\delta_{qs}}{\sqrt{6}}$ & $-g_{u}\frac{2\delta_{qu}-\delta_{qd}-\delta_{qs}}{\sqrt{6}}$\tabularnewline
\cline{1-4} \cline{6-9} 
\end{tabular}
\end{table}

\begin{eqnarray}
c_{\pm}^{(2)}\left(x,\xi\right) & = & \left(\int dz\frac{\phi_{2}(z)}{z}\right)\frac{8\pi i}{9}\frac{\alpha_{s}\left(\mu_{R}^{2}\right)f_{M}}{Q}\frac{1}{x\pm\xi\mp i0}\left(1+\frac{\alpha_{s}\left(\mu_{r}^{2}\right)}{2\pi}T^{(1)}\left(\frac{x\pm\xi}{2\xi},\, z\right)\right).\label{eq:c2}
\end{eqnarray}
where $\phi_{2}(z)$ is the twist-2 $\pi-$or $K$-meson distribution
amplitude (DA) defined as~\cite{Kopeliovich:2011rv} 
\begin{eqnarray}
\phi_{2}\left(z\right) & = & \frac{1}{if_{M}\sqrt{2}}\int\frac{du}{2\pi}e^{i(z-0.5)u}\left\langle 0\left|\bar{\psi}\left(-\frac{u}{2}n\right)\hat{n}\gamma_{5}\psi\left(\frac{u}{2}n\right)\right|\pi(q)\right\rangle .\label{eq:DA2p}
\end{eqnarray}

The function $T^{(1)}\left(v,\, z\right)$ in~(\ref{eq:c2}) encodes
NLO corrections to the coefficient function. As was explained in~\cite{Belitsky:2001nq,Ivanov:2004zv,Diehl:2007hd}
it is related by analytical continuation to the loop correction to
$\bar{q}q$ scattering, and was evaluated and analyzed in detail in
the context of NLO studies of the pion form factor (see~\cite{Braaten:1987yy,Melic:1998qr}
for details and historical discussion). Explicitly, it is given by
\begin{align}
T^{(1)}\left(v,\, z\right) & =\frac{1}{2vz}\left[\frac{4}{3}\left([3+\ln(v\, z)]\,\ln\left(\frac{Q^{2}}{\mu_{F}^{2}}\right)+\frac{1}{2}\ln^{2}\left(v\, z\right)+3\ln(v\, z)-\frac{\ln\bar{v}}{2\bar{v}}-\frac{\ln\bar{z}}{2\bar{z}}-\frac{14}{3}\right)\right.\label{eq:T1}\\
\nonumber \\
 & +\beta_{0}\left(\frac{5}{3}-\ln(v\, z)-\ln\left(\frac{Q^{2}}{\mu_{R}^{2}}\right)\right)\nonumber \\
\nonumber \\
 & -\frac{1}{6}\left(2\frac{\bar{v}\, v^{2}+\bar{z}\, z^{2}}{(v-z)^{3}}\left[{\rm Li}_{2}(\bar{z})-{\rm Li}_{2}(\bar{v})+{\rm Li}_{2}(v)-{\rm Li}_{2}(z)+\ln\bar{v}\,\ln z-\ln\bar{z}\,\ln v\right]\right.\nonumber \\
 & +2\frac{v+z-2v\, z}{(v-z)^{2}}\ln\left(\bar{v}\bar{z}\right)+2\left[{\rm Li}_{2}(\bar{z})+{\rm Li}_{2}(\bar{v})-{\rm Li}_{2}(z)-{\rm Li}_{2}(v)+\ln\bar{v}\,\ln z+\ln\bar{z}\,\ln v\right]\nonumber \\
 & +\left.\left.4\frac{v\, z\,\ln(v\, z)}{(v-z)^{2}}-4\ln\bar{v}\,\ln\bar{z}-\frac{20}{3}\right)\right],\nonumber 
\end{align}
where $\beta_{0}=\frac{11}{3}N_{c}-\frac{2}{3}N_{f}$, ${\rm Li}_{2}(z)$
is the dilogarithm function, and $\mu_{R}$ and $\mu_{F}$ are the renormalization
and factorization scales respectively~%
\footnote{For the sake of simplicity, we follow~\cite{Diehl:2007hd} and assume
that the factorization scale $\mu_{F}$ is the same for both the generalized
parton distribution and the pion distribution amplitude.%
}. The correction~$T^{(1)}\left(v,\, z\right)$ for small $v\approx0$
($x=\pm\xi\mp i0$) has an asymptotic behavior $\sim\ln^{2}v$, which
signals that a collinear approximation might be not valid near this
point. To regularize the singularity, we may follow~\cite{Goloskokov:2009ia}
and introduce a small transverse momentum $l_{\perp}$ of the quark
inside a meson. Effectively, this corresponds to the introduction of
a small infrared regularization in the region $v\sim l_{\perp}^{2}/Q^{2}$,
a vanishingly small quantity in the Bjorken limit. However, a full evaluation
of $T^{(1)}\left(v,\, z\right)$ beyond collinear approximation (taking into
account all higher twist corrections) presents a challenging problem.
Another possibility was suggested in~\cite{Diehl:2007hd}, and corresponds
to the absorption of the singular term by selecting a low renormalization
scale $\mu_{R}^{2}\sim z\, v\, Q^{2}$. Near the points $x\approx\pm\xi$
the redefined scale $\mu_{R}$ drops to very small values, where nonperturbative
effects become relevant. Only in the Bjorken limit ($Q^{2}\to\infty$)
we may expect that details of regularization become irrelevant.

\section{GPD and DA parametrizations}

\label{sec:Parametrizations} For the leading twist DA $\phi_{2\pi}(x)$,
the currently available data on the meson photoproduction form factor
$F_{\pi\gamma\gamma}\left(Q^{2}\right)$ are compatible with the asymptotic
form $\phi_{as}(z)=6\sqrt{2}f_{\pi}z(1-z)$, with a typical uncertainty
in the minus-first moment of the order of $\sim10\%$ (see e.g.~\cite{Pimikov:2012nm,Bakulev:2012nh}
and reviews in~\cite{Brodsky:2011xx,Brodsky:2011yv}).

More than a dozen different parametrizations of GPDs have been
proposed in the literature~\cite{Diehl:2000xz,Goloskokov:2008ib,Radyushkin:1997ki,Kumericki:2011rz,Guidal:2010de,Polyakov:2008aa,Polyakov:2002wz,Freund:2002qf,Goldstein:2013gra}.
While we neither endorse nor refute any of them, for the sake of concreteness
we use the parametrization~\cite{Goloskokov:2006hr,Goloskokov:2007nt,Goloskokov:2008ib},
which succeeded to describe HERA~\cite{Aaron:2009xp} and JLAB~\cite{Goloskokov:2006hr,Goloskokov:2007nt,Goloskokov:2008ib}
data on electroproduction of different mesons, so it should provide
a reasonable description of $\nu$DVMP. The parametrization is based
on the Radyushkin's double distribution ansatz. It assumes additivity
of the valence and sea parts of the GPDs, 
\begin{equation}
H(x,\xi,t)=H_{val}(x,\xi,t)+H_{sea}(x,\xi,t),
\end{equation}
which are defined as 
\begin{eqnarray}
H_{val}^{q} & = & \int_{|\alpha|+|\beta|\le1}d\beta d\alpha\delta\left(\beta-x+\alpha\xi\right)\,\frac{3\theta(\beta)\left((1-|\beta|)^{2}-\alpha^{2}\right)}{4(1-|\beta|)^{3}}q_{val}(\beta)e^{\left(b_{i}-\alpha_{i}\ln|\beta|\right)t},\label{eq:H_val}\\
H_{sea}^{q} & = & \int_{|\alpha|+|\beta|\le1}d\beta d\alpha\delta\left(\beta-x+\alpha\xi\right)\,\frac{3\, sgn(\beta)\left((1-|\beta|)^{2}-\alpha^{2}\right)^{2}}{8(1-|\beta|)^{5}}q_{sea}(\beta)e^{\left(b_{i}-\alpha_{i}\ln|\beta|\right)t},\label{eq:H_sea}
\end{eqnarray}
and $q_{val}$ and $q_{sea}$ are the ordinary valence and sea components
of the PDFs. The coefficients $b_{i}$, $\alpha_{i}$, as well as the
parametrization of the input PDFs $q(x),\,\Delta q(x)$ and pseudo-PDFs
$e(x),\,\tilde{e}(x)$ (which correspond to the forward limit of the
GPDs $E,\,\tilde{E}$), are discussed in~\cite{Goloskokov:2006hr,Goloskokov:2007nt,Goloskokov:2008ib}.
The unpolarized PDFs $q(x)$ are adjusted to reproduce the CTEQ PDFs
in the limited range $4\lesssim Q^{2}\lesssim40$~GeV$^{2}$. Notice
that in this model the sea is flavor symmetric for asymptotically
large $Q^{2}$, 
\begin{equation}
H_{sea}^{u}=H_{sea}^{d}=\kappa\left(Q^{2}\right)H_{sea}^{s},\label{eq:SeaFlavourSymmetry}
\end{equation}
where 
\[
\kappa\left(Q^{2}\right)=1+\frac{0.68}{1+0.52\ln\left(Q^{2}/Q_{0}^{2}\right)},\quad Q_{0}^{2}=4\,{\rm GeV^{2}}.
\]

The equality of the sea components of the light quarks in~(\ref{eq:SeaFlavourSymmetry})
should be considered only as a rough approximation, since in the forward
limit the inequality $\bar{d}\not=\bar{u}$ was firmly established
by the E866/NuSea experiment~\cite{Hawker:1998ty}. For this reason,
predictions done with this parametrization of GPDs for the $p\rightleftarrows n$
transitions in the region $x_{Bj}\in(0.1...0.3)$ might slightly underestimate
the data.

\section{Numerical results and discussion}

\label{sec:Results}

In this section we would like to present numerical results for the
next-to-leading order corrections to pion production using the Kroll-Goloskokov
parametrization of GPDs~\cite{Goloskokov:2006hr,Goloskokov:2007nt,Goloskokov:2008ib,Goloskokov:2009ia},
briefly discussed in section~\ref{sec:Parametrizations}. Due
to poor statistics of the neutrino-induced processes, it is challenging
to measure the differential cross-section $d\sigma/dx_{B}dt\, dQ^{2}$,
so we will restrict ourselves to the cross-section $d^{2}\sigma/dx_{B}dQ^{2}$.
Using for reference the kinematics of MINERvA experiment~\cite{Drakoulakos:2004gn},
we assume that the average energy of the neutrino beam is 6~GeV.
The predicted cross-section change only mildly when we smear out the
cross-section with a realistic spectrum.

\begin{figure}
\includegraphics{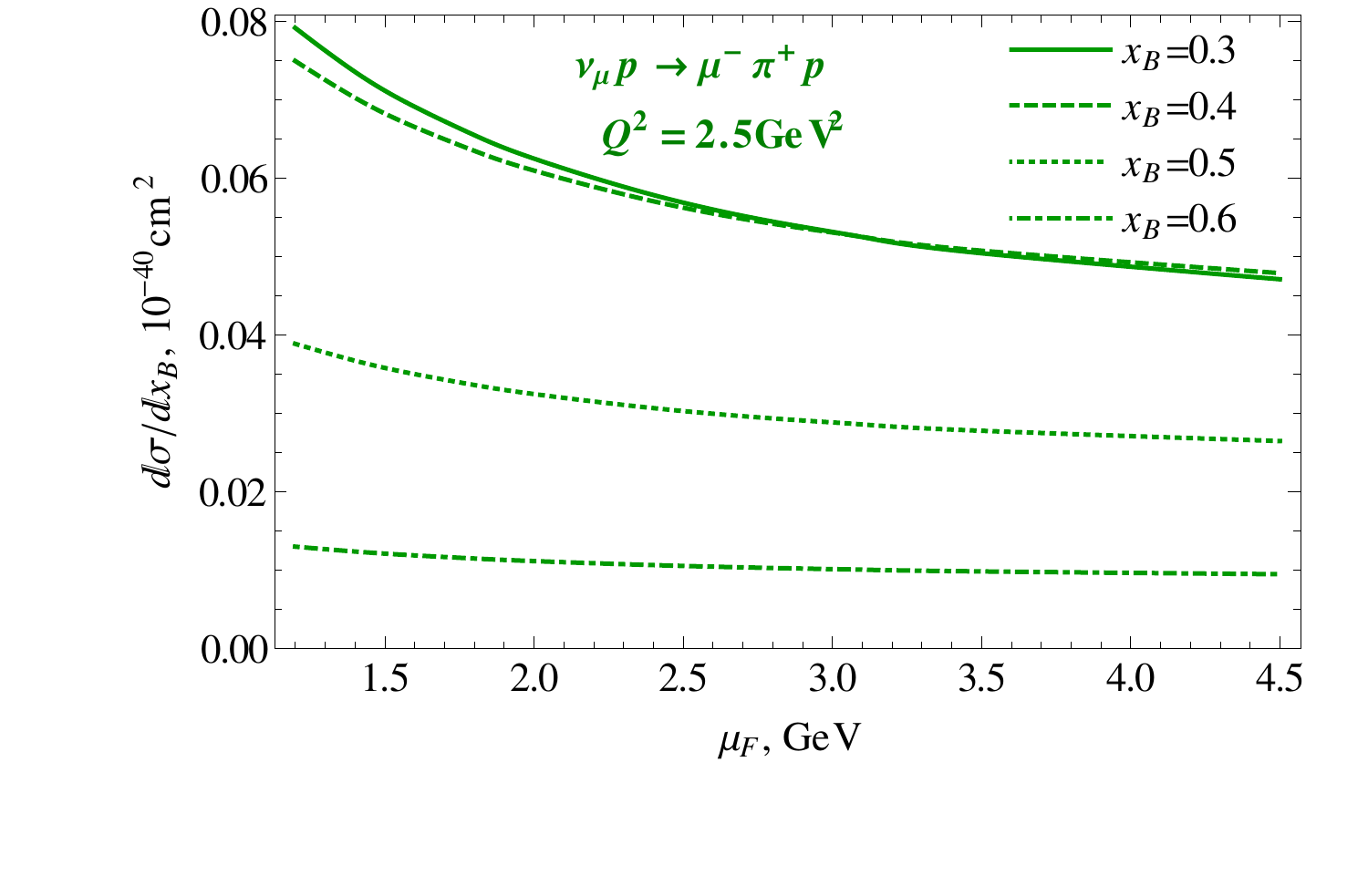}

\protect\caption{\label{fig:DVMP-pions-off}(color online) Factorization scale dependence
of the charged current $\pi^{+}$ production for the process $\nu p\to\mu^{-}\pi^{+}p$
for several values of $x_{B}$. Similar dependence is observed for
all other processes.}
\end{figure}

We would like to start a discussion about the dependence on the factorization
scale $\mu_{F}$, which separates hard and soft physics. As we can
see from Figure~\ref{fig:DVMP-pions-off}, the results become
independent on the factorization scale $\mu_{F}$ only at a sufficiently
large $\mu_{F}\gtrsim5$~GeV. Though the choice of factorization
scale $\mu_{F}$ is arbitrary, a choice of a value significantly different
from the virtuality $Q$ could lead to large logarithms in higher order
corrections. As was suggested in~\cite{Belitsky:2001nq,Ivanov:2004zv,Diehl:2007hd},
varying the scale in the range $\mu_{F}\in\left(Q/2,\,2Q\right)$,
we can roughly estimate the error due to omitted higher order loop
contributions.

\begin{figure}
\includegraphics[scale=0.75]{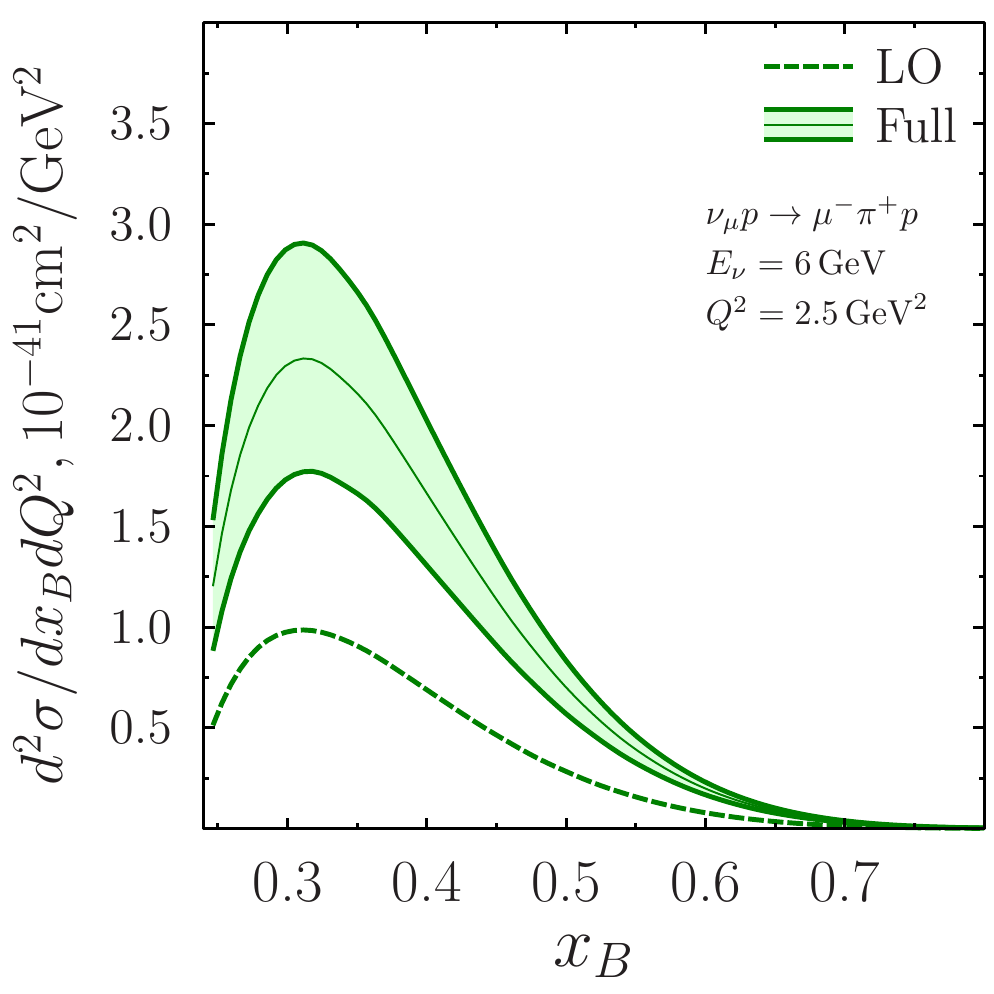}\includegraphics[scale=0.75]{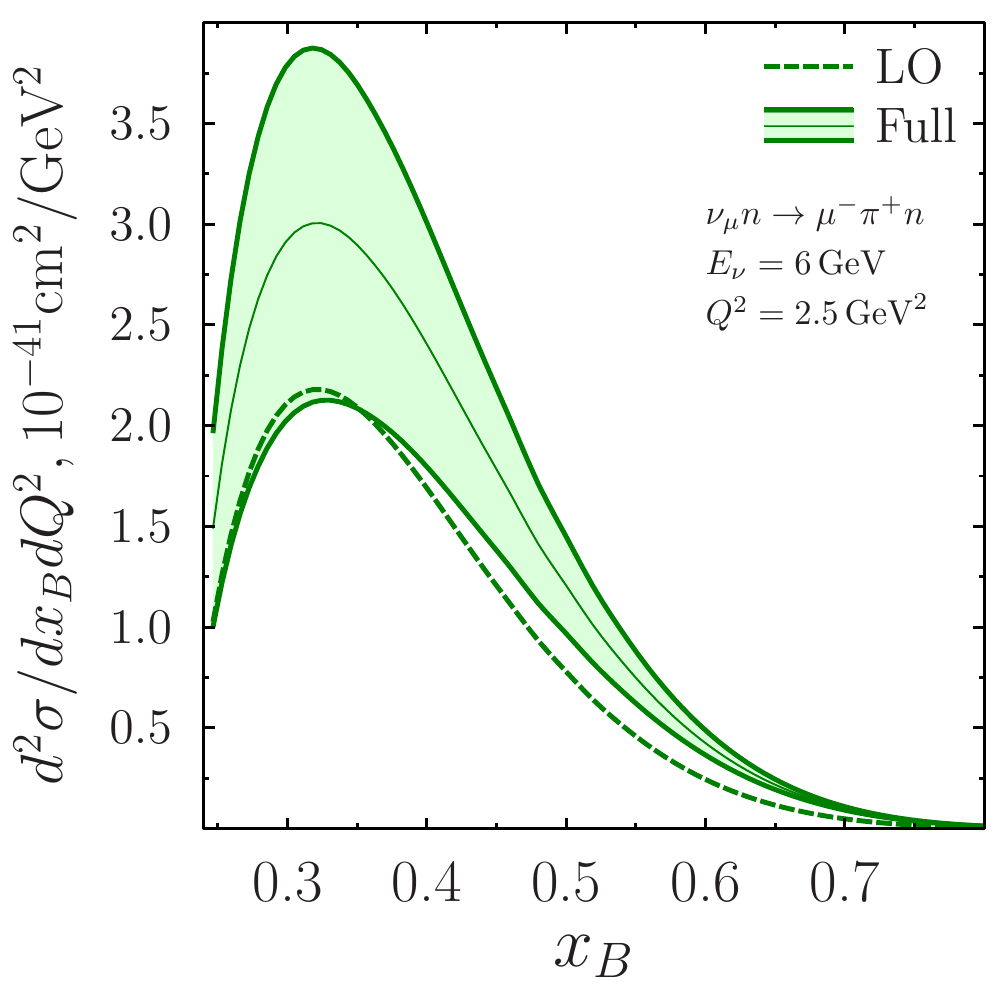}\\
\includegraphics[scale=0.75]{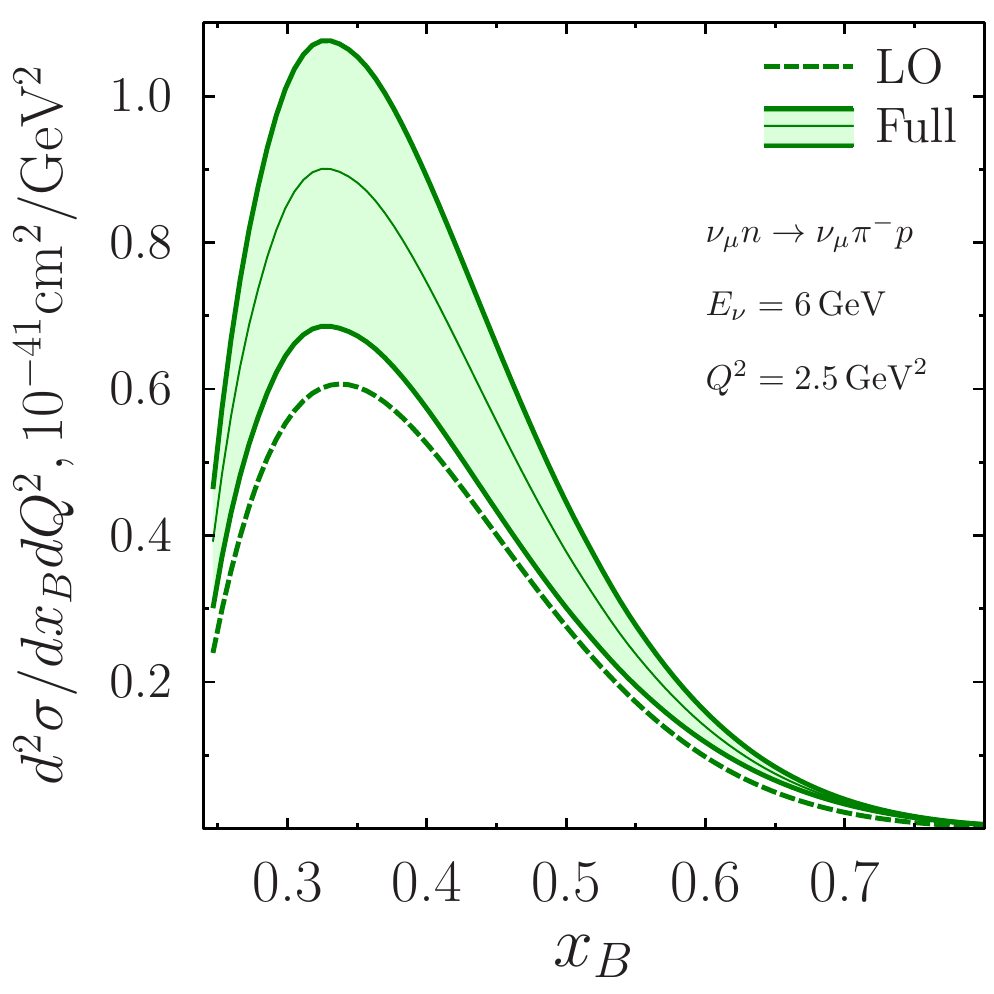}\includegraphics[scale=0.75]{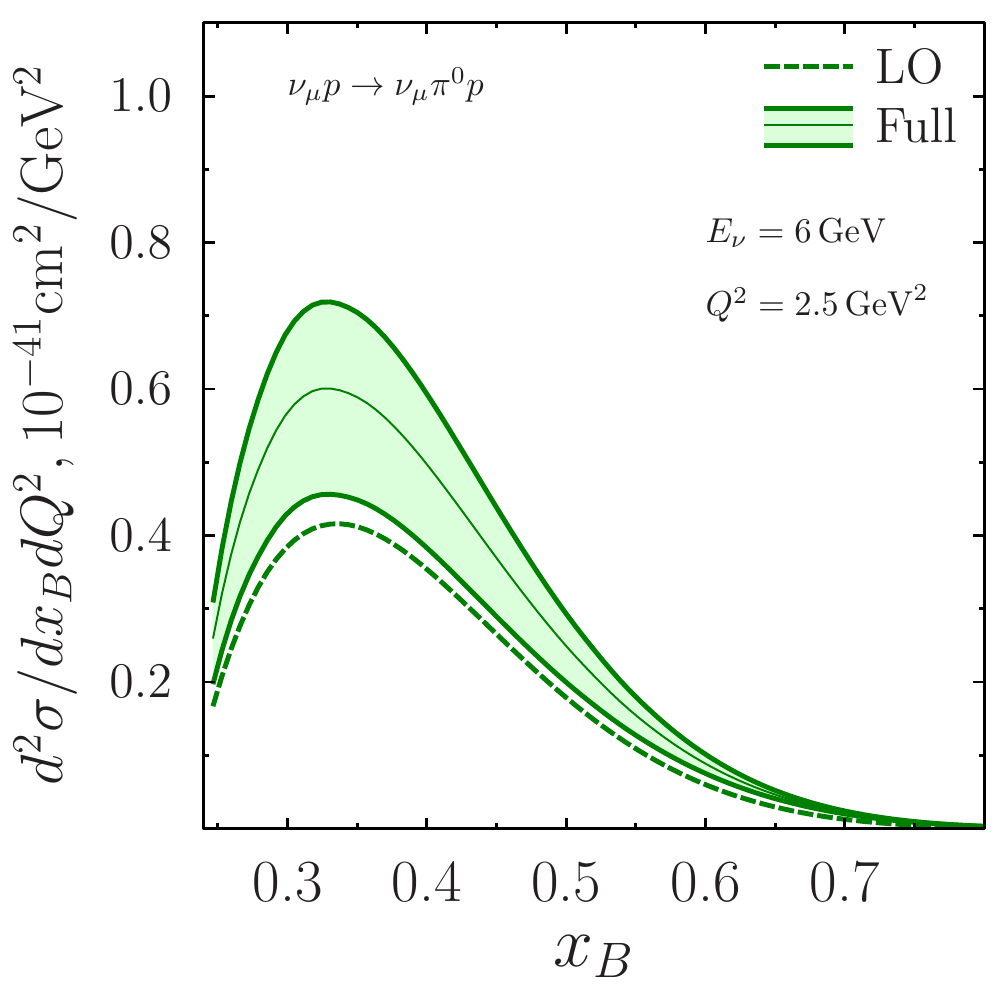}\\
\includegraphics[scale=0.75]{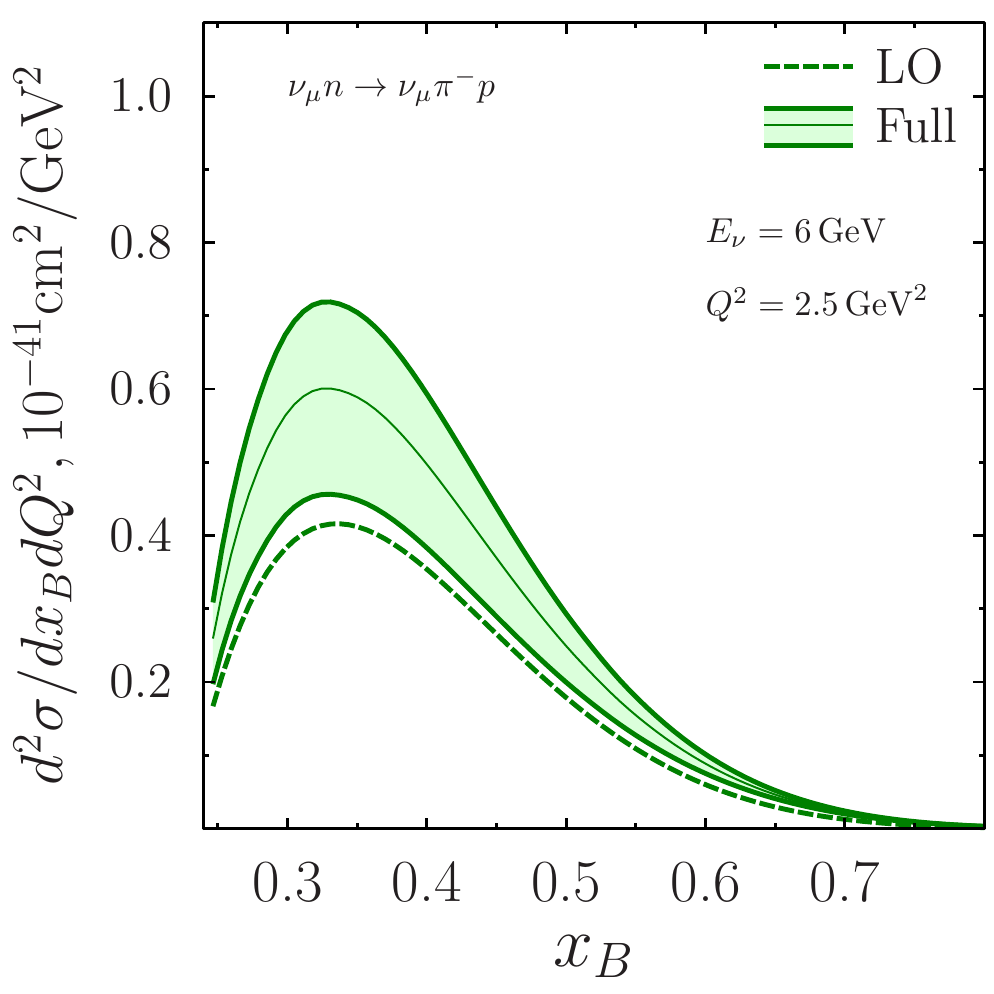}\includegraphics[scale=0.75]{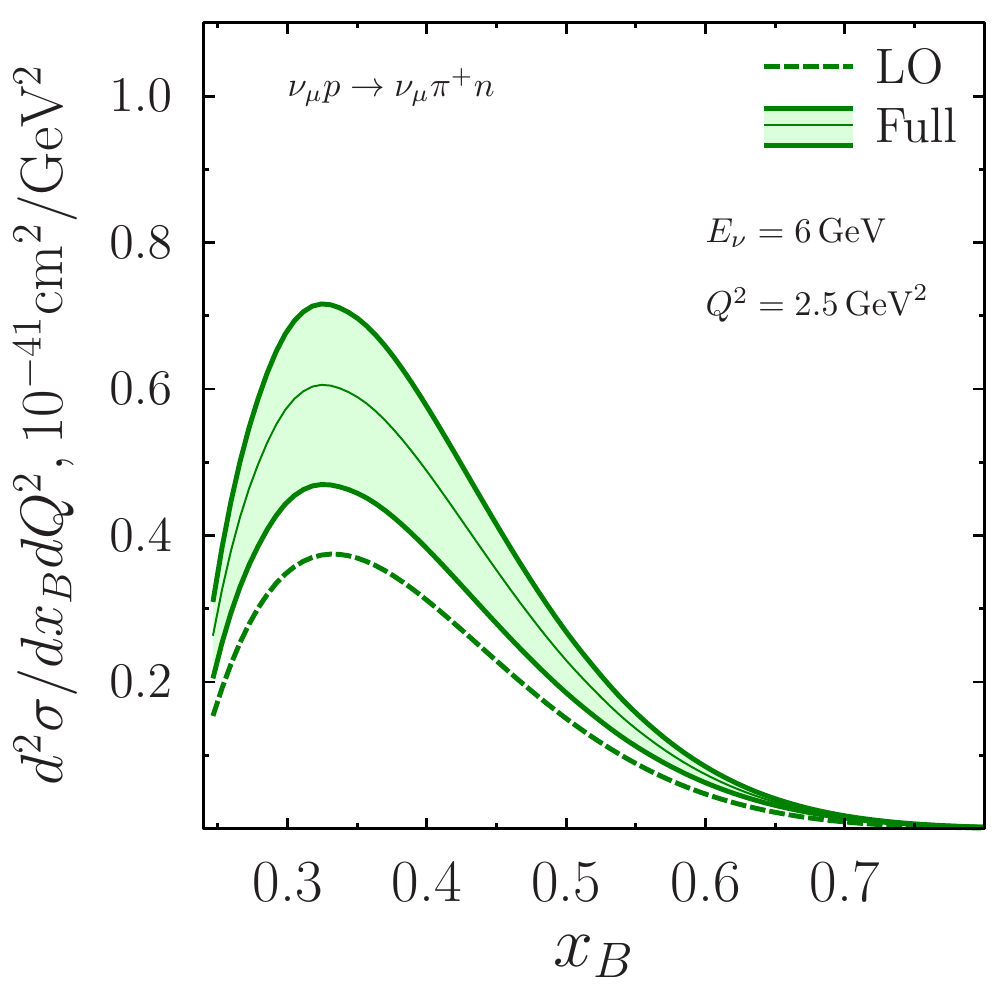}\\

\protect\caption{\label{fig:DVMP-pions}(color online) Pion production on nucleons,
with neutral and charged currents at fixed energy neutrino beam
($E_{\nu}\approx6$~GeV). The dashed line stands for the leading order
evaluation, whereas the solid line surrounded by green error bands (marked
as ``Full'') stands for the full result with NLO corrections. The
width of the band represents the uncertainty due to the factorization
scale choice~$\mu_{F}\in\left(Q/2,\,2Q\right)$, as explained in the text. }
\end{figure}

In Figure~\ref{fig:DVMP-pions} we show the predictions for
the differential cross-section $d\sigma/dx_{B}dQ^{2}$ for charged
and neutral pion production in several channels. For all cross-sections,
at fixed neutrino energy $E_{\nu}$ and virtuality $Q^{2}$, we have
a similar bump-like shape, which is explained by a competition of
two factors. For small $x_{B}\sim Q^{2}/2m_{N}E_{\nu}$ the elasticity
$y$ defined in~(\ref{eq:elasticity}) approaches one, which causes
a suppression due to a prefactor in~(\ref{eq:sigma_def}). In the
opposite limit, a suppression $\sim(1-x)^{n}$ is controlled by the
implemented parametrization of GPD. As we can see from a comparison
of the leading order (dashed lines) and the full results (solid line surrounded
by the green band), the next-to-leading order corrections are sizable
and increase the full cross-section by $\sim$50\%. In the charged
current case, $\pi^{+}$ production on protons, the cross-section is even
larger: as we explained in~\cite{Kopeliovich:2012dr}, in the leading
order there is a partial cancellation of the $s$-channel and $u$-channel
handbag contributions, which leads to a twice smaller cross-section
compared to the same process on neutrons in leading order. However,
for the next-to-leading order such cancellation no longer occurs,
which explains the elevated NLO correction to the charged current
$\pi^{+}$ production on protons.

\begin{figure}
\includegraphics[scale=0.75]{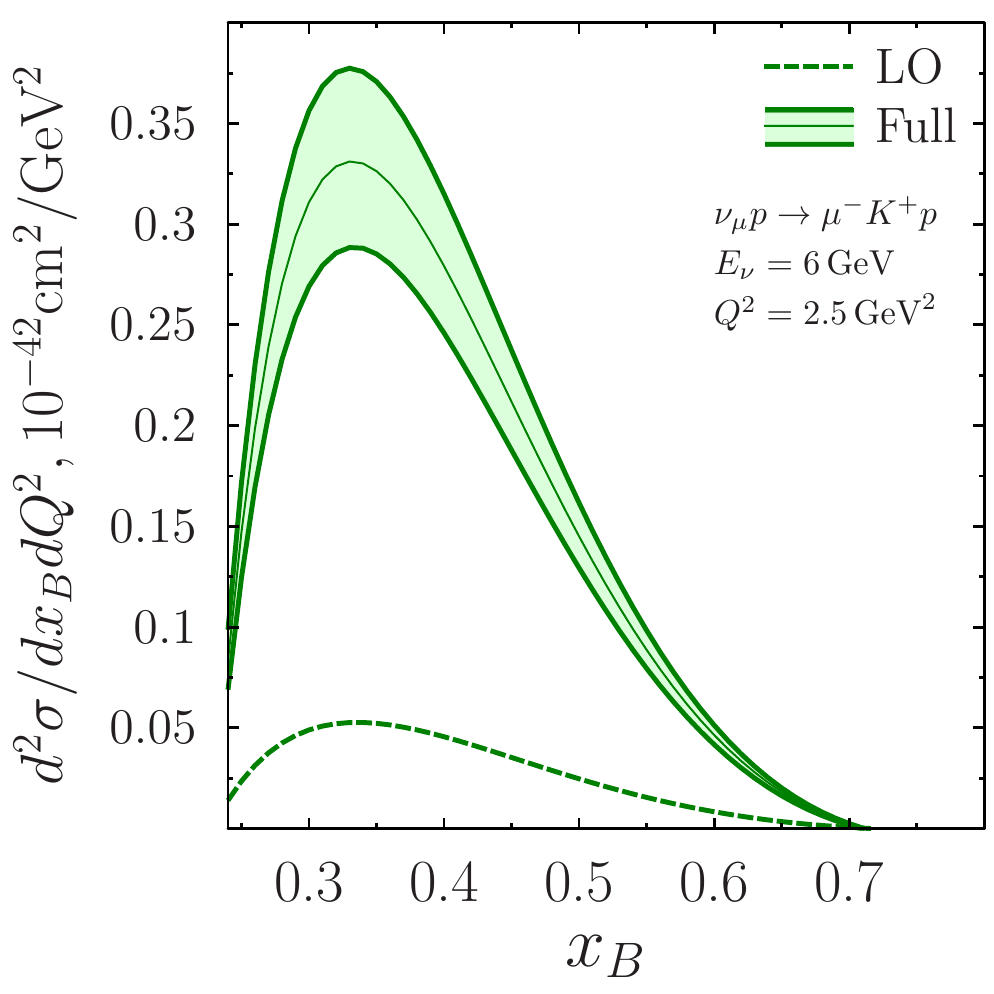}\includegraphics[scale=0.75]{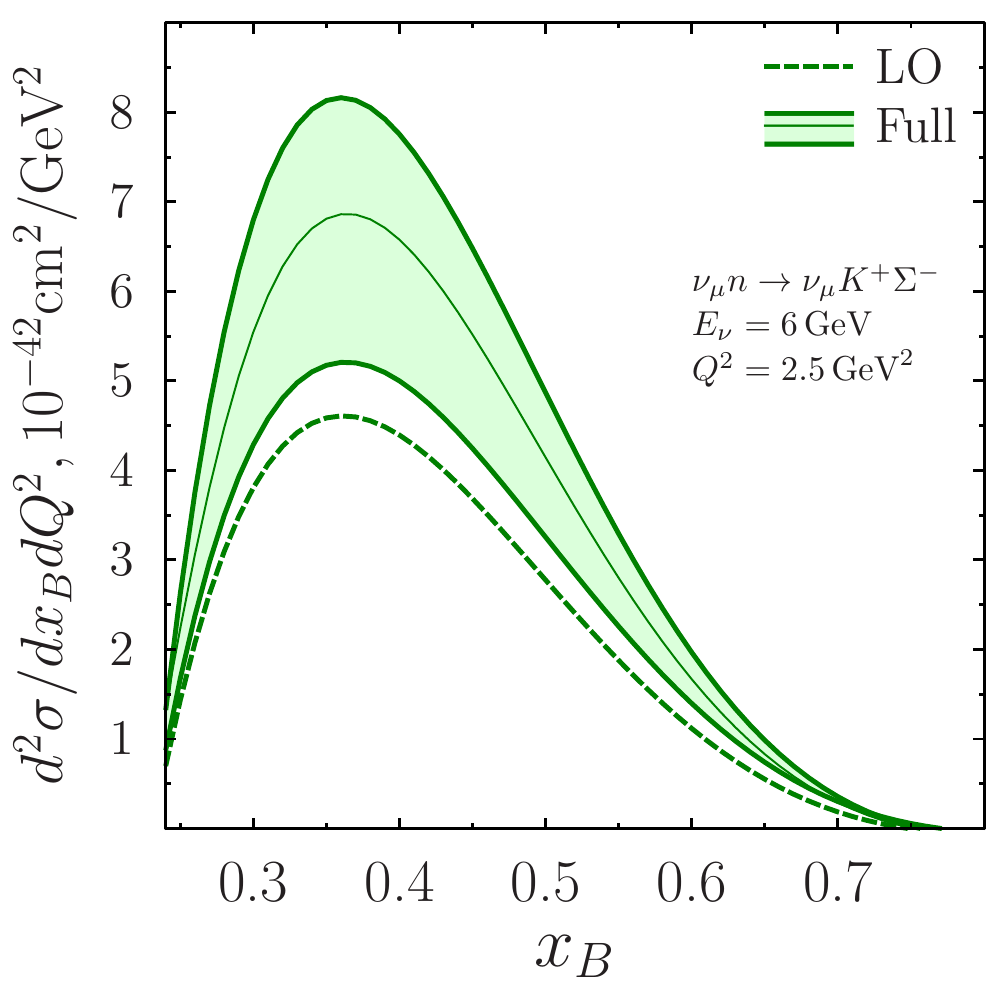}\\
\includegraphics[scale=0.75]{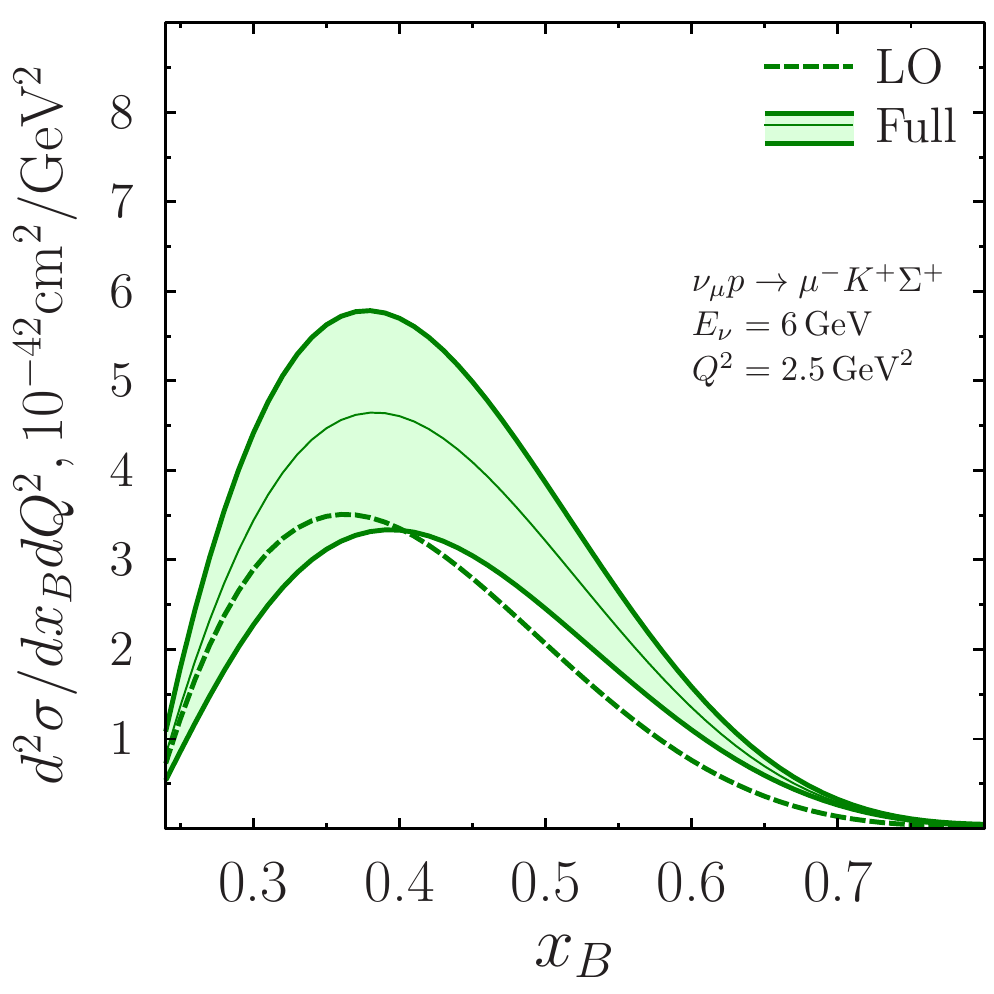}\includegraphics[scale=0.75]{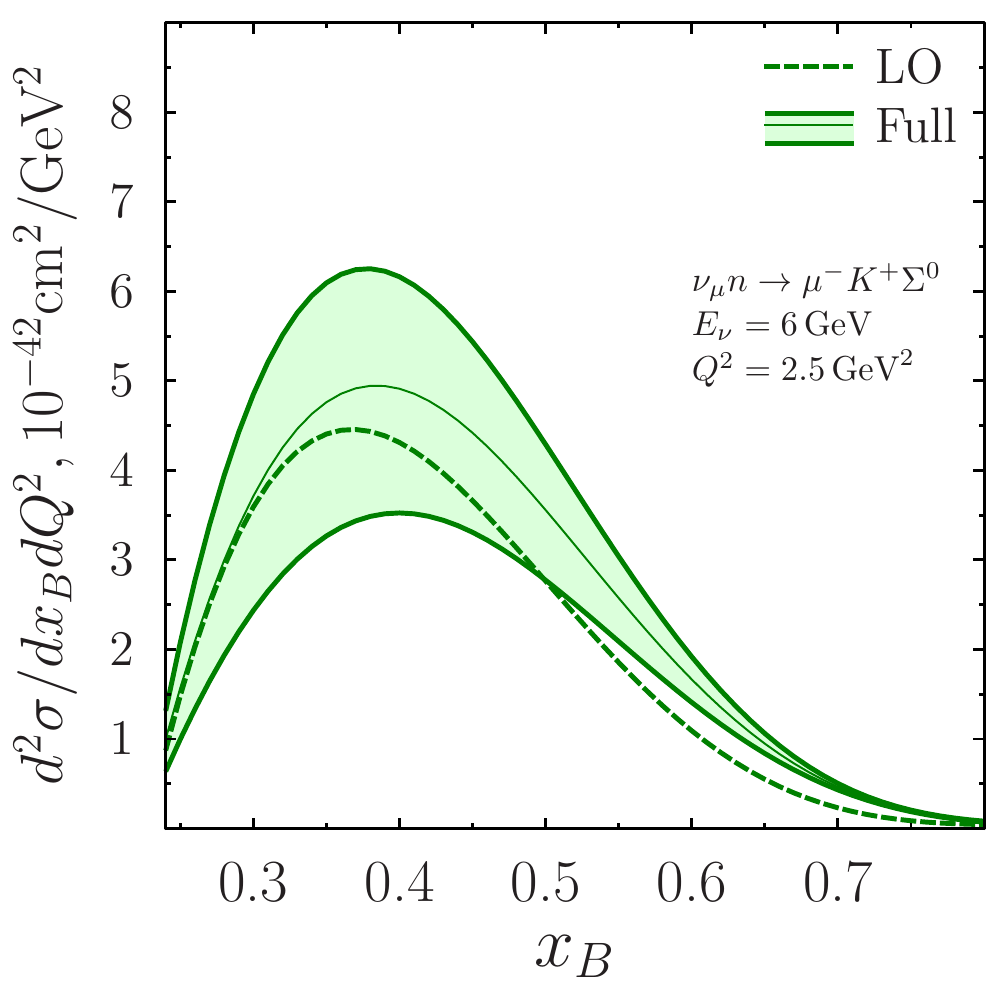}\\
\includegraphics[scale=0.75]{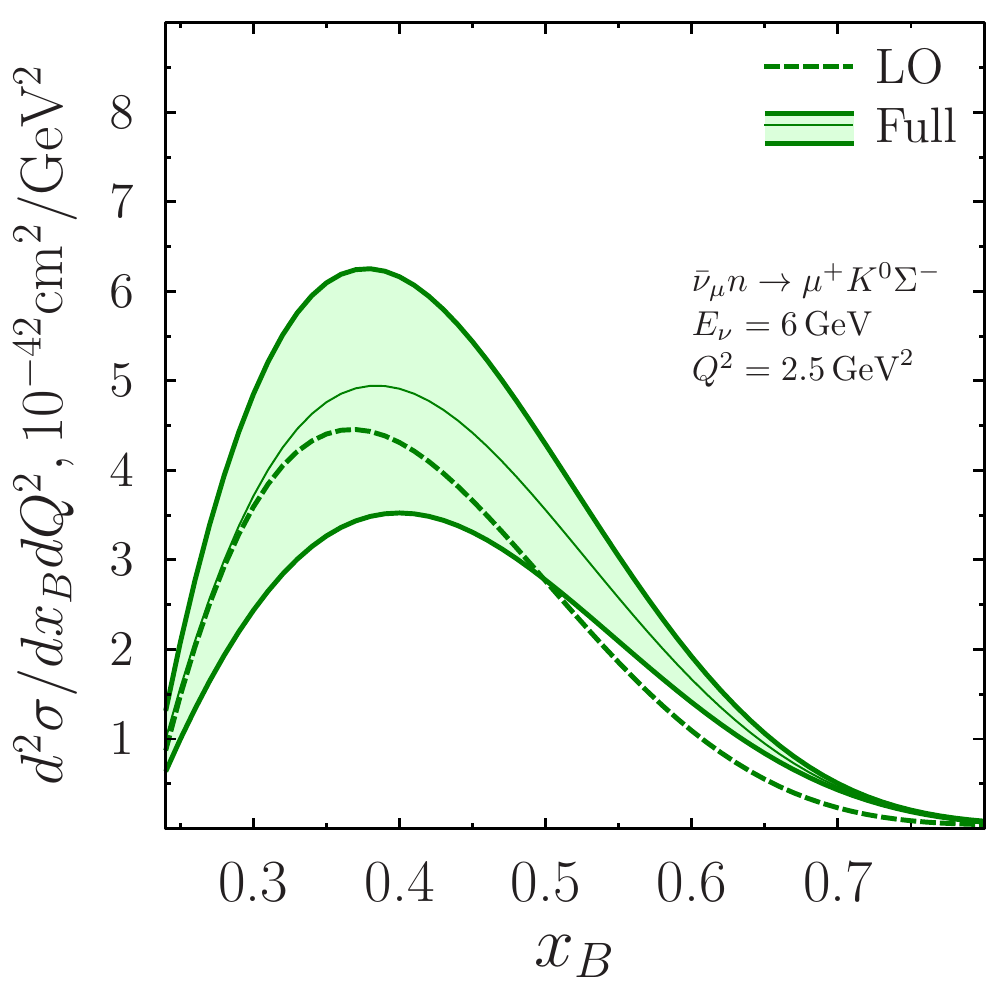}\includegraphics[scale=0.75]{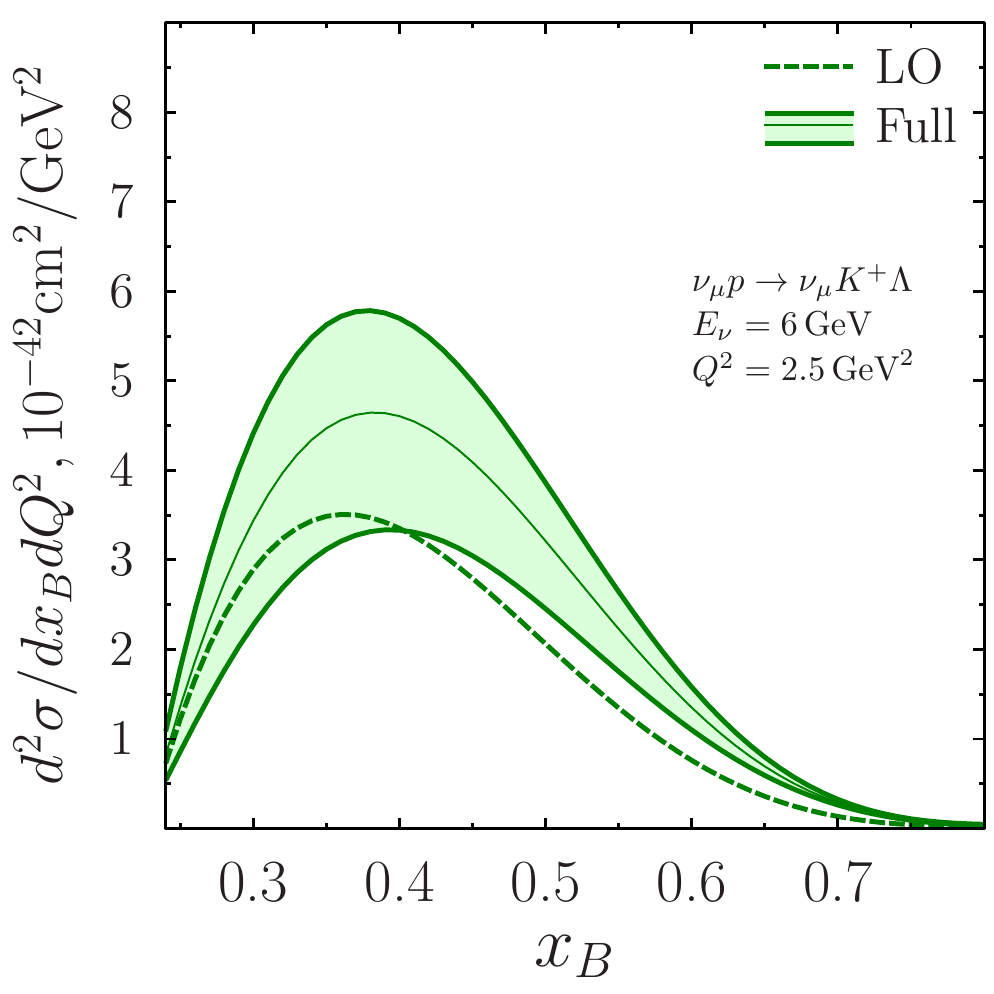}\\

\protect\caption{\label{fig:DVMP-kaons}(color online) Selected neutral and charged
current mediated kaon production cross-sections for fixed energy neutrino
beam ($E_{\nu}\approx6$~GeV).  The dashed line stands for the leading
order evaluation, whereas the solid line surrounded by green error bands
(marked as ``Full'') stands for the full result, which takes into account 
NLO corrections. The width of the band represents the uncertainty due
to the factorization scale choice~$\mu_{F}\in\left(Q/2,\,2Q\right)$,
as explained in the text. }
\end{figure}

Similarly, for the case of kaon production (see Figure~\ref{fig:DVMP-kaons}),
we observe that corrections are large. From the upper left plot we
can see that Cabibbo suppressed ($\Delta S=1$) $K^{+}$-production
on the proton has extremely small cross-section, beyond the reach
of ongoing and forthcoming experiments, and for this reason we do not
consider other Cabibbo suppressed channels. The Cabbibo-allowed~($\Delta S=0$)
processes have an order of magnitude larger cross-sections and potentially
could be used to test the poorly known strange quark GPD. 

\section{Conclusions}

In this paper we estimated the contributions of the next-to-leading
order corrections to pion and kaon production in neutrino-nucleus
collisions. We found that these corrections increase the full cross-section
by a factor of 1.5-2, and for this reason are important in the analysis of
generalized parton distributions from the data. The NLO coefficient
functions near the points $x\pm\xi$ have logarithmic behavior, which
suggests that higher twist corrections in the NLO might be important,
especially in the imaginary part. As was discussed in~\cite{Kopeliovich:2013ae},
such corrections generate the azimuthal angle dependence, which could
be used to assess the size of these harmonics. However, at this moment
a systematic evaluation of NLO corrections at twist 3 presents a challenging
problem.

Qualitatively, our findings agree with large NLO corrections to meson
\emph{electro}production~\cite{Belitsky:2001nq,Ivanov:2004zv,Diehl:2007hd},
deeply virtual Compton scattering~\cite{Freund:2001hd,Belitsky:1999sg,Belitsky:1997rh}
and timelike Compton scattering~\cite{Pire:2011st}, expected in electron-induced
processes. In view of this result, a future analysis of the next-to-next-to-leading
order corrections is desirable. Our results are relevant for the analysis
of pion and kaon production in the \textsc{Minerva }experiment
at FERMILAB as well as the planned~Muon Collider/Neutrino Factory~\cite{Gallardo:1996aa,Ankenbrandt:1999as,Alsharoa:2002wu}. 

A code for the evaluation of the cross-sections, taking into account NLO
corrections and employing various GPD models is available on demand.

\section*{Acknowledgments}

This research was partially supported by Proyecto Basal FB 0821 (Chile),
the Fondecyt (Chile) grants 1140390 and 1140377, CONICYT (Chile) grant
PIA ACT1413. Powered@NLHPC: This research was partially supported
by the supercomputing infrastructure of the NLHPC (ECM-02). Also,
we thank Yuri Ivanov for technical support of the USM HPC cluster
where part of evaluations were done.

\appendix


\begin{thebibliography}{10}
 \bibitem{Ji:1998xh} X.~D.~Ji and J.~Osborne, Phys.\ Rev.\ D
 \textbf{58} (1998) 094018 {[}arXiv:hep-ph/9801260{]}.
 
 \bibitem{Collins:1998be} J.~C.~Collins and A.~Freund, Phys.\ Rev.\ D
 \textbf{59}, 074009 (1999).
 
 \bibitem{Mueller:1998fv} D.~Mueller, D.~Robaschik, B.~Geyer, F.~M.~Dittes
 and J.~Horejsi, Fortsch.\ Phys.\ \textbf{42}, 101 (1994) {[}arXiv:hep-ph/9812448{]}.
 
 \bibitem{Ji:1996nm} X.~D.~Ji, Phys.\ Rev.\ D \textbf{55}, 7114
 (1997).
 
 \bibitem{Ji:1998pc} X.~D.~Ji, J.\ Phys.\ G \textbf{24}, 1181
 (1998) {[}arXiv:hep-ph/9807358{]}.
 
 \bibitem{Radyushkin:1996nd} A.~V.~Radyushkin, Phys.\ Lett.\ B
 \textbf{380}, 417 (1996) {[}arXiv:hep-ph/9604317{]}.
 
 \bibitem{Radyushkin:1997ki} A.~V.~Radyushkin, Phys.\ Rev.\ D
 \textbf{56}, 5524 (1997).
 
 \bibitem{Radyushkin:2000uy} A.~V.~Radyushkin, arXiv:hep-ph/0101225.
 
 \bibitem{Collins:1996fb} J.~C.~Collins, L.~Frankfurt and M.~Strikman,
 Phys.\ Rev.\ D \textbf{56}, 2982 (1997).
 
 \bibitem{Brodsky:1994kf} S.~J.~Brodsky, L.~Frankfurt, J.~F.~Gunion,
 A.~H.~Mueller and M.~Strikman, Phys.\ Rev.\ D \textbf{50}, 3134
 (1994).
 
 \bibitem{Goeke:2001tz} K.~Goeke, M.~V.~Polyakov and M.~Vanderhaeghen,
 Prog.\ Part.\ Nucl.\ Phys.\ \textbf{47}, 401 (2001) {[}arXiv:hep-ph/0106012{]}.
 
 \bibitem{Diehl:2000xz} M.~Diehl, T.~Feldmann, R.~Jakob and P.~Kroll,
 Nucl.\ Phys.\ B \textbf{596}, 33 (2001) {[}Erratum-ibid.\ B \textbf{605},
 647 (2001){]} {[}arXiv:hep-ph/0009255{]}.
 
 \bibitem{Belitsky:2001ns} A.~V.~Belitsky, D.~Mueller and A.~Kirchner,
 Nucl.\ Phys.\ B \textbf{629}, 323 (2002) {[}arXiv:hep-ph/0112108{]}.
 
 \bibitem{Diehl:2003ny} M.~Diehl, Phys.\ Rept.\ \textbf{388}, 41
 (2003) {[}arXiv:hep-ph/0307382{]}.
 
 \bibitem{Belitsky:2005qn} A.~V.~Belitsky and A.~V.~Radyushkin,
 Phys.\ Rept.\ \textbf{418}, 1 (2005) {[}arXiv:hep-ph/0504030{]}.
 
 \bibitem{Kubarovsky:2011zz}V.~Kubarovsky {[}CLAS Collaboration{]},
 Nucl.~Phys.~Proc.~Suppl.~\textbf{219-220}, 118 (2011).
 
 \bibitem{Ahmad:2008hp} S.~Ahmad, G.~R.~Goldstein and S.~Liuti,
 Phys.\ Rev.\ D \textbf{79} (2009) 054014 {[}arXiv:0805.3568 {[}hep-ph{]}{]}.
 
 \bibitem{Goloskokov:2009ia}S.~V.~Goloskokov and P.~Kroll, Eur.~Phys.~J.~C
 \textbf{65}, 137 (2010) {[}arXiv:0906.0460 {[}hep-ph{]}{]}.
 
 \bibitem{Goloskokov:2011rd}S.~V.~Goloskokov and P.~Kroll, Eur.~Phys.~J.~A
 \textbf{47}, 112 (2011) {[}arXiv:1106.4897 {[}hep-ph{]}{]}.
 
 \bibitem{Goldstein:2012az}G.~R.~Goldstein, J.~O.~G.~Hernandez
 and S.~Liuti, arXiv:1201.6088 {[}hep-ph{]}.
 
 \bibitem{Kopeliovich:2012dr}B.~Z.~Kopeliovich, I.~Schmidt and
 M.~Siddikov, Phys. Rev. D \textbf{86} (2012), 113018 {[}arXiv:1210.4825
 {[}hep-ph{]}{]}.
 
 \bibitem{Drakoulakos:2004gn}D.~Drakoulakos \emph{et al.} {[}Minerva
 Collaboration{]}, 
  hep-ex/0405002.
 
 \bibitem{Kopeliovich:2014pea}B.~Z.~Kopeliovich, I.~Schmidt and
 M.~Siddikov, Phys.~Rev.~D \textbf{89}, no. 5, 053001 (2014) {[}arXiv:1401.1547
 {[}hep-ph{]}{]}.
 
 \bibitem{Goldstein:2009in}G.~R.~Goldstein, O.~G.~Hernandez, S.~Liuti
 and T.~McAskill, AIP Conf.~Proc.~\textbf{1222}, 248 (2010) {[}arXiv:0911.0455
 {[}hep-ph{]}{]}.
 
 \bibitem{Frankfurt:1999fp}L.~L.~Frankfurt, P.~V.~Pobylitsa, M.~V.~Polyakov
 and M.~Strikman, 
  Phys.~Rev.~D \textbf{60} (1999) 014010 {[}hep-ph/9901429{]}.
 
 \bibitem{Pire:2015iza}B.~Pire and L. Szymanowski, Phys.~Rev.~Lett.~\textbf{115}
 (2015), 092001 {[}arXiv:1505.00917 {[}hep-ph{]}{]}.
 
 \bibitem{Pire:2015vxa}B.~Pire and L.~Szymanowski, Acta Phys.~Polon.~Supp.~
 \textbf{8}, 883 (2015) {[}arXiv:1510.01869 {[}hep-ph{]}{]}.
 
 \bibitem{Pire:2016jtr}B.~Pire, L.~Szymanowski and J.~Wagner, EPJ Web Conf.~\textbf{112}, 01018 (2016) {[}arXiv:1601.07666 {[}hep-ph{]}{]}.
 
 \bibitem{Kopeliovich:2013ae} B.~Z.~Kopeliovich, I.~Schmidt and
 M.~Siddikov, Phys.\ Rev.\ D \textbf{87}, 033008 (2013) {[}arXiv:1301.7014
 {[}hep-ph{]}{]}.
 
 \bibitem{Schmidt:2015nka}I.~Schmidt and M.~Siddikov, Phys.~Rev.~D
 \textbf{91}, no. 7, 073002 (2015) {[}arXiv:1501.04306 {[}hep-ph{]}{]}.
 
 \bibitem{Vanderhaeghen:1998uc}M.~Vanderhaeghen, P.~A.~M.~Guichon
 and M.~Guidal, 
  Phys.~Rev.~Lett.~\textbf{80}, 5064 (1998).
 
 \bibitem{Mankiewicz:1998kg}L.~Mankiewicz, G.~Piller and A.~Radyushkin,
 \textbf{10}, 307 (1999) {[}hep-ph/9812467{]}.
 
 \bibitem{Goloskokov:2006hr}S.~V.~Goloskokov and P.~Kroll, 
  Eur.~Phys.~J.~C \textbf{50}, 829 (2007) {[}hep-ph/0611290{]}.
 
 \bibitem{Goloskokov:2007nt}S.~V.~Goloskokov and P.~Kroll, 
  Eur.~Phys.~J.~C \textbf{53}, 367 (2008) {[}arXiv:0708.3569 {[}hep-ph{]}{]}.
 
 \bibitem{Goloskokov:2008ib}S.~V. Goloskokov and P.~Kroll, Eur.~Phys.~J.~C
 \textbf{59} (2009) 809 {[}arXiv:0809.4126 {[}hep-ph{]}{]}.
 
 \bibitem{Kopeliovich:2011rv}B.~Z.~Kopeliovich, Iván Schmidt and
 M.~Siddikov, Nucl.~Phys.~A \textbf{918}, 41 (2013) {[}arXiv:1108.5654
 {[}hep-ph{]}{]}.
 
 \bibitem{Belitsky:2001nq}A.~V.~Belitsky and D.~Mueller, Phys.~Lett.~B
 \textbf{513}, 349 (2001) {[}hep-ph/0105046{]}.
 
 \bibitem{Ivanov:2004zv}D.~Y.~Ivanov, L.~Szymanowski and G.~Krasnikov,
 JETP Lett.~\textbf{80}, 226 (2004) {[}Pisma Zh.~Eksp.~Teor.~Fiz.~\textbf{80},
 255 (2004){]} Erratum: {[}JETP Lett.~\textbf{101}, no. 12, 844 (2015){]}
 , {[}hep-ph/0407207{]}.
 
 \bibitem{Diehl:2007hd}M.~Diehl and W.~Kugler, Eur.~Phys.~J.~C
 \textbf{52}, 933 (2007) {[}arXiv:0708.1121 {[}hep-ph{]}{]}.
 
 \bibitem{Braaten:1987yy}E.~Braaten and S.~M.~Tse, Phys.~Rev.~D
 \textbf{35}, 2255 (1987).
 
 \bibitem{Melic:1998qr}B.~Melic, B.~Nizic and K.\textasciitilde{}Passek,
 Phys.~Rev.~\textbf{60}, 074004 (1999) {[}hep-ph/9802204{]}.
 
 \bibitem{Pimikov:2012nm}A.~V.~Pimikov, A.~P.~Bakulev, S.~V.~Mikhailov
 and N.~G.~Stefanis, arXiv:1208.4754 {[}hep-ph{]}.
 
 \bibitem{Bakulev:2012nh}A.~P.~Bakulev, S.~V.~Mikhailov, A.~V.~Pimikov
 and N.~G.~Stefanis, Phys.~Rev.~D \textbf{86} (2012) 031501 {[}arXiv:1205.3770
 {[}hep-ph{]}{]}.
 
 \bibitem{Brodsky:2011xx}S.~J. Brodsky, F.~-G.~Cao and G.~F.~de
 Teramond, 
  Phys.~Rev.~D \textbf{84}, 075012 (2011) {[}arXiv:1105.3999 {[}hep-ph{]}{]}.
 
 \bibitem{Brodsky:2011yv}S.~J.~Brodsky, F.~-G.~Cao and G.~F.~de
 Teramond, 
  Phys.~Rev.~D \textbf{84}, 033001 (2011) {[}arXiv:1104.3364 {[}hep-ph{]}{]}.
 
 \bibitem{Kumericki:2011rz}K.~Kumericki, D.~Muller and A.~Schafer,
 JHEP \textbf{1107}, 073 (2011) {[}arXiv:1106.2808 {[}hep-ph{]}{]}.
 
 \bibitem{Guidal:2010de}M.~Guidal, 
  Phys.~Lett.~B \textbf{693}, 17 (2010) {[}arXiv:1005.4922 {[}hep-ph{]}{]}.
 
 \bibitem{Polyakov:2008aa}M.~V.~Polyakov and K.~M.~Semenov-Tian-Shansky,
 Eur.~Phys.~J.~A \textbf{40}, 181 (2009) {[}arXiv:0811.2901 {[}hep-ph{]}{]}.
 
 \bibitem{Polyakov:2002wz}M.~V.~Polyakov and A.~G.~Shuvaev, 
  hep-ph/0207153.
 
 \bibitem{Freund:2002qf}A.~Freund, M.~McDermott and M.~Strikman,
 Phys.~Rev.~D \textbf{67}, 036001 (2003) {[}hep-ph/0208160{]}.
 
 \bibitem{Goldstein:2013gra}G.~R.~Goldstein, J.~O.~G.~Hernandez
 and S.~Liuti, arXiv:1311.0483 {[}hep-ph{]}.
 
 \bibitem{Aaron:2009xp}F.~D.~Aaron \emph{et al.} {[}H1 Collaboration{]},
 JHEP \textbf{1005} (2010) 032 {[}arXiv:0910.5831 {[}hep-ex{]}{]}.Phys.~Rev.~D
 \textbf{82}, 033001 (2010) {[}arXiv:1004.5484 {[}hep-ph{]}{]}.
 
 \bibitem{Hawker:1998ty}E.~A.~Hawker \emph{et al.} {[}FNAL E866/NuSea
 Collaboration{]}, Phys.~Rev.~Lett.~\textbf{80} (1998) 3715 {[}hep-ex/9803011{]}.
 
 \bibitem{Freund:2001hd}A.~Freund and M.~McDermott, Eur.~Phys.~J.~C
 \textbf{23}, 651 (2002) {[}hep-ph/0111472{]}.
 
 \bibitem{Belitsky:1999sg}A.~V.~Belitsky, D.~Mueller, L.~Niedermeier
 and A.~Schafer, Phys.~Lett.~B \textbf{474}, 163 (2000) {[}hep-ph/9908337{]}.
 
 \bibitem{Belitsky:1997rh}A.~V.~Belitsky and D.~Mueller, Phys.~Lett.~B
 \textbf{417}, 129 (1998) {[}hep-ph/9709379{]}.
 
 \bibitem{Pire:2011st}B.~Pire, L.~Szymanowski and J.~Wagner, Phys.~Rev.~D
 \textbf{83}, 034009 (2011) {[}arXiv:1101.0555 {[}hep-ph{]}{]}.
 
 \bibitem{Gallardo:1996aa}J. C. Gallardo, R.~B. Palmer, A.~V.~Tollestrup,
 A. M.~Sessler, A.~N.~Skrinsky, C.~Ankenbrandt, S.~Geer and J.~Griffin
 \emph{et al.}, 
  eConf C \textbf{960625} (1996) R4.
 
 \bibitem{Ankenbrandt:1999as}C.~M.~Ankenbrandt, M.~Atac, B.~Autin,
 V.~I.~Balbekov, V.~D.~Barger, O.~Benary, J.~S.~Berg and M.~S.~Berger
 \emph{et al.}, 
  Phys.~Rev.~ST Accel.~Beams \textbf{2} (1999) 081001 {[}physics/9901022{]}.
 
 \bibitem{Alsharoa:2002wu}M.~M.~Alsharoa \emph{et al.} {[}Muon Collider/Neutrino
 Factory Collaboration{]}, 
  Phys.~Rev.~ST Accel.~Beams \textbf{6} (2003) 081001 {[}hep-ex/0207031{]}.
 
\end{thebibliography}
 \end{document}